\documentclass[aps,reprint,showpacs,amsmath,amssymb,prl,superscriptaddress]{revtex4-2}
\usepackage{amsmath,amssymb,latexsym,epsfig,graphics,epsf}

\begin{document}

\title{Breakdown of material-time scaling in the microscopic dynamics of metallic glasses}

\date{\today}

\author{Till Böhmer}
\email{till.boehmer@dlr.de}
\affiliation{Institute of Frontier Materials on Earth and in Space, German Aerospace Center, D-51147 Cologne, Germany}
\author{Jie Shen}
\affiliation{Institut Néel, Université Grenoble Alpes and Centre National de la Recherche
Scientifique, Grenoble, France}
\affiliation{European Synchrotron Radiation Facility, Grenoble, France}
\author{Jan Philipp Gabriel}
\affiliation{Institute of Frontier Materials on Earth and in Space, German Aerospace Center, D-51147 Cologne, Germany}
\author{Rolf Zeißler}
\affiliation{Institute for Condensed Matter Physics, Technical University of Darmstadt, D-64289 Darmstadt, Germany}
\author{Federico Zontone}
\affiliation{European Synchrotron Radiation Facility, Grenoble, France}
\author{Fan Yang}
\affiliation{Institute of Frontier Materials on Earth and in Space, German Aerospace Center, D-51147 Cologne, Germany}
\author{Eloi Pineda}
\affiliation{Department of Physics, Institute of Energy Technologies, Universitat Politècnica de Catalunya—BarcelonaTech, Barcelona, Spain}
\author{Beatrice Ruta}
\email{beatrice.ruta@neel.cnrs.fr}
\affiliation{Institut Néel, Université Grenoble Alpes and Centre National de la Recherche Scientifique, Grenoble, France}
\affiliation{European Synchrotron Radiation Facility, Grenoble, France}
\author{Thomas Blochowicz}
\email{thomas.blochowicz@pkm.tu-darmstadt.de}
\affiliation{Institute for Condensed Matter Physics, Technical University of Darmstadt, D-64289 Darmstadt, Germany}

\begin{abstract}
Physical aging in glasses is often described using the concept of material time, which assumes the underlying relaxation mechanism to remain the same as in equilibrium, but with a rate that changes during aging. Whether material time also captures the microscopic dynamics of metallic glasses, which differ qualitatively between the glass and the equilibrium supercooled liquid, is unclear. Using simultaneous time-resolved X-ray diffraction and X-ray photon-correlation spectroscopy, we follow the full equilibration of a metallic glass after small temperature jumps starting from equilibrium. The time evolution of the first sharp diffraction peak position exhibits the asymmetry between temperature up- and down-jumps characteristic of material-time aging and is quantitatively described by the Tool--Narayanaswamy--Moynihan model, which is based on this concept. By contrast, the microscopic density-density autocorrelation functions strongly violate material-time scaling. Compressed-exponential relaxation emerges even after temperature changes as small as 4\,K, irrespective of the direction of the temperature jump, and continuously evolves into a stretched-exponential relaxation shape following the equilibration of the average structure. This behavior is consistent with a transient stress-mediated contribution associated with volumetric frustration that disappears upon equilibration. Our results show that a material time can describe the evolution of average structural properties even when the corresponding microscopic dynamics does not obey material-time scaling.
\end{abstract}

\maketitle

Glass formation occurs when structural relaxation becomes too slow for a supercooled liquid to remain in equilibrium on the experimental timescale. The resulting glass is therefore not in an equilibrium thermodynamic state, but in a kinetically arrested one whose properties depend on its preparation history. With time, considering only structural rearrangements without affecting the chemical composition, the glass continues to evolve towards the equilibrium supercooled liquid state through the process defined as \textit{physical aging}~\cite{Struik1978,Scherer1986,Hodge1995}. Physical aging is accompanied by changes in macroscopic properties such as density~\cite{Kovacs1963} and mechanical response~\cite{Struik1978,Ketkaew2018,Ketkaew2018b}, and affects the stability and lifetime of glassy materials~\cite{Hodge1995}. Understanding and predicting how these properties evolve with age therefore remains an important problem in glass physics~\cite{Hodge1994,McKenna2017}.

A central idea in the description of physical aging is that the relevant relaxation mechanism remains the same as in the equilibrium supercooled liquid, but proceeds at a rate that changes continuously during aging. This idea has been formalized in terms of a framework that extends relations valid in thermal equilibrium, such as linear response~\cite{Narayanaswami1971} or time-reversal symmetry~\cite{Boehmer2024}, to aging glasses by replacing the ordinary time $t$ with an internal clock, referred to as the \textit{material time} $\xi$. Its rate, $\gamma(t)=\mathrm{d}\xi/\mathrm{d}t$, is non-uniform and evolves continuously during aging to reflect the changing relaxation rate~\cite{Narayanaswami1971,Hodge1994,Hecksher2010,Hecksher2015,Dyre2015,Boehmer2024,Zhao2025}.

The concept of material time forms the basis of phenomenological aging models such as the Tool--Narayanaswamy--Moynihan (TNM) formalism~\cite{Tool1946,Narayanaswami1971,Debolt1976} and naturally accounts for characteristic features of physical aging, including the asymmetry of approach following temperature up- and down-jumps~\cite{Hodge1994}. For jumps towards the same final temperature, a down-jump typically approaches equilibrium faster and with a more stretched relaxation shape than an up-jump~\cite{Scherer1986, McKenna2017, Dyre2015}. In the material-time picture, this asymmetry follows from the opposite evolution of the internal clock rate: after a down-jump, the initially fast dynamics slow during aging, whereas after an up-jump the initially slow dynamics accelerate~\cite{Riechers2022,Hecksher2010,Hecksher2015}.

More recently, the material-time concept has been extended from describing the evolution of macroscopic observables, to microscopic fluctuations probed through two-time correlation functions $C(t_1,t_2)$. If the microscopic dynamics can be described in terms of a material time, the two-time correlation function depends only on the elapsed material time, such that
\begin{equation}
    C(t_1,t_2) = \phi\bigl[\xi(t_2)-\xi(t_1) \bigr],
\end{equation}
where $\phi$ is an age-independent relaxation function~\cite{Dyre2015,Boehmer2024}.

For three times $t_1<t_2<t_3$, this implies the triangular relation~\cite{Cugliandolo1994,Dyre2015,Boehmer2024,Amari2026}
\begin{equation}
    C(t_1,t_3) = F\bigl[ C(t_1,t_2),C(t_2,t_3) \bigr],
    \label{eq:triangular-relation}
\end{equation}
where $F$ depends only on the two correlation values and is independent of the age $t_1$~\cite{Boehmer2024}. A violation of Eq.~(\ref{eq:triangular-relation}) therefore directly falsifies a material-time description of the measured dynamics.

Direct tests of material-time predictions, beyond fits to phenomenological aging models such as TNM, have so far been performed predominantly for molecular glasses, using dielectric, mechanical, and light-scattering observables in controlled temperature-jump experiments close to the glass transition~\cite{Hecksher2010,Hecksher2015,Riechers2022,Boehmer2024}. These experiments typically start from an equilibrated supercooled liquid, apply a small temperature perturbation, and subsequently follow the complete relaxation towards the new equilibrium state under isothermal conditions.

In this regard, metallic glasses (MGs) provide an interesting case because their microscopic relaxation observed in the non-equilibrium glass differs qualitatively from that of the equilibrium supercooled liquid. X-ray photon correlation spectroscopy (XPCS) experiments have repeatedly revealed compressed-exponential intermediate scattering functions (ISF) in MGs, commonly described by
\begin{equation}
    f(t_1,t_2) \propto \exp \left[ -\left(\frac{t_2-t_1}{\tau}\right)^\beta\,\right]
\end{equation}
with stretching-exponent $\beta>1$~\cite{Ruta2012,Ruta2013,Giordano2016,Ruta2017,Amini2021,Frey2025,Shen2026}. This contrasts with the stretched-exponential ($\beta<1$) structural relaxation typically observed in supercooled liquids~\cite{ruta2020,Neuber2022,Boehmer2025}. Compressed relaxation in MGs has been associated with stress-mediated collective atomic rearrangements~\cite{Ruta2012,Ruta2017,Frey2025,Shen2026}. Recent work has further related it to medium-range order (MRO) structural correlations and suggested that it may constitute an additional non-equilibrium relaxation mechanism beyond the equilibrium structural relaxation~\cite{Frey2025,Shen2026}. Whether this apparent transition from stretched- to compressed-exponential relaxation at the glass transition can still be accounted for by a material-time variable is not clear. Such behavior is, however, theoretically possible, since a time-dependent clock rate can itself modify the shape of a correlation function when expressed in laboratory time~\cite{Boehmer2024,Amari2026}.



Previous XPCS studies of aging in MGs have predominantly employed substantial temperature changes or sequential thermal protocols, often driving the system far from equilibrium. Even when measurements were performed close to $T_g$, the state was commonly reached by heating from much lower temperatures, so that its distance from equilibrium at the measurement temperature was not directly controlled~\cite{Ruta2012,Ruta2013,Ruta2017,Giordano2016,Evenson2015}. Such protocols do not allow to distinguish an intrinsic breakdown of material-time scaling in MGs from deviations caused by a large departure from equilibrium, since material-time descriptions are known to break down for sufficiently large temperature changes even in other glass formers~\cite{Henot2024,Amari2026}. It therefore remains an open question whether aging in MGs conforms to a material-time description when tested under well-controlled conditions. Moreover, it is not clear whether compressed-exponential relaxation emerges even following small, controlled temperature jumps from equilibrium. If it does, such a temperature protocol allows us to resolve how the compressed dynamics develop and evolve as the system returns to equilibrium.

To address these questions, we apply a controlled temperature-jump strategy to a MG. Starting from an equilibrated supercooled liquid state at temperature $T_0$, we impose small but fast temperature up- and down-jumps of varying magnitude $\Delta T$ and follow the evolution of both the microscopic dynamics and the average structure until equilibrium at the final temperature $T_\infty$ is recovered. We find that compressed-exponential relaxation emerges even after these small perturbations and evolves continuously towards the stretched-exponential equilibrium shape as the structure equilibrates. At the same time, the microscopic two-time correlations violate the triangular relation and are therefore incompatible with a material-time description. By contrast, the evolution of the average structure remains consistent with material-time aging, as it is quantitatively described by the TNM model. These results show that the evolution of the average structure can remain consistent with a material-time description even when the microscopic dynamics during equilibration is not.


\begin{figure*}[t!]
    \centering
       \includegraphics[width=0.99\textwidth]{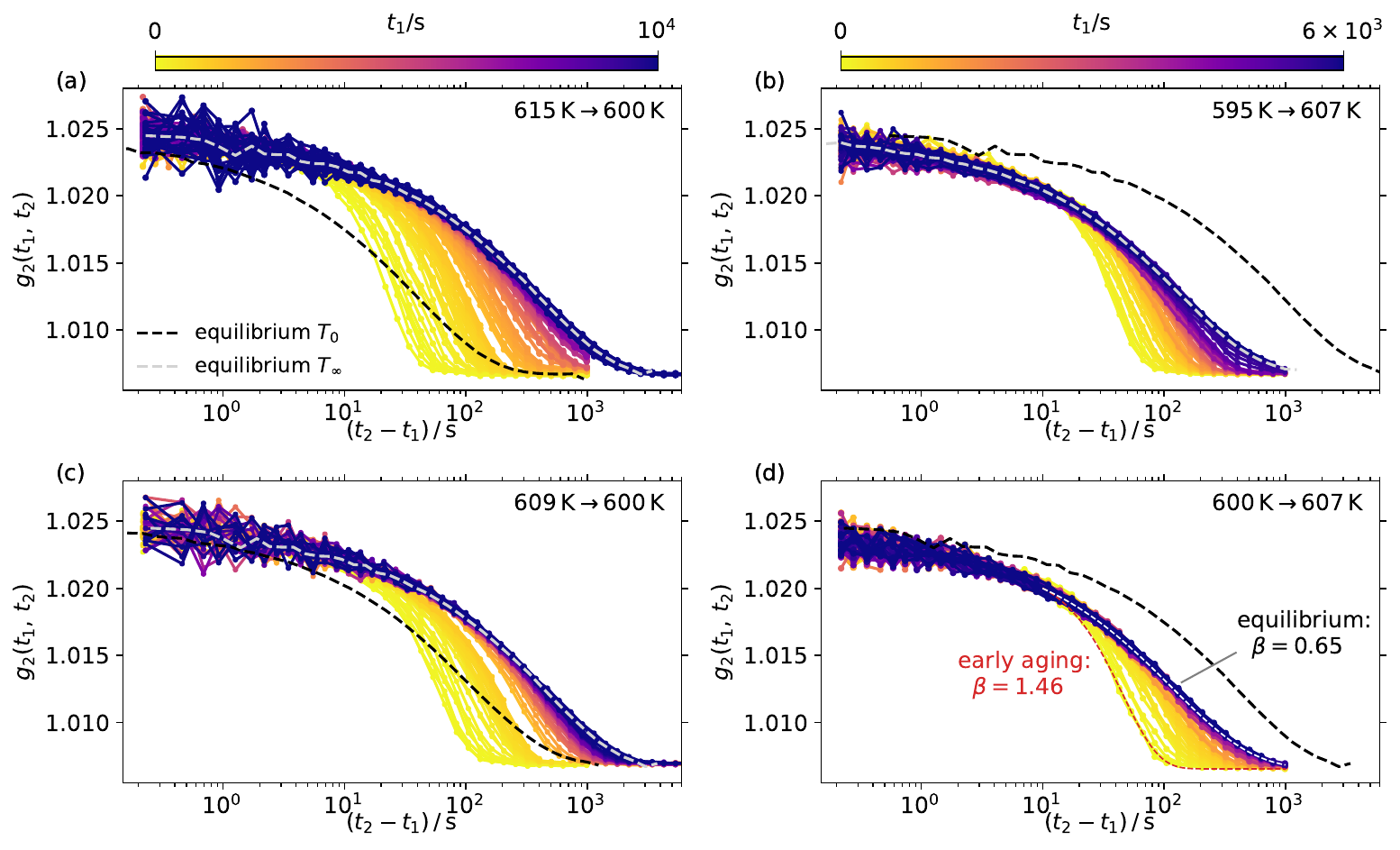}
    \caption{\label{fig1}
    Time-resolved intensity autocorrelation functions $g_2(q,t_1,t_2)$ during aging measured at $q=2.61\,\mathrm{\AA}^{-1}$ plotted as a function of the lag time $t_2-t_1$ following different temperature jumps. Down-jumps are shown in panels (a) and (c), and up-jumps in panels (b) and (d). The color indicates the waiting time $t_1$, with $t_1=0$ defined as the time at which the sample reaches $T_\infty$. Equilibrium autocorrelation functions at the initial temperature $T_0$ and final temperature $T_\infty$ are shown as black and gray dashed lines, respectively. Panel (d) shows two representative fits of the autocorrelation functions using Eq.~(\ref{equ:KWW}), illustrating the evolution of the stretching exponent from $\beta>1$ at early waiting times towards the equilibrium value $\beta=0.65$ during aging.
    }
\end{figure*}

\section*{Results}

To follow physical aging simultaneously at the level of the average structure and the microscopic dynamics, we perform time-resolved X-ray diffraction (XRD) and X-ray photon correlation spectroscopy (XPCS) on the MG Zr$_{46.8}$Ti$_{8.2}$Cu$_{7.5}$Ni$_{10}$Be$_{27.5}$ (Vit4) at the ID10-COH beamline of the ESRF. Experimental details are given in the Materials and Methods section.

XRD provides the time- and $q$-dependent scattering intensity $I(q,t)$. We characterize the evolution of the average structure by the position $q_{\mathrm{FSDP}}(t)$ of its first sharp diffraction peak (FSDP). Simultaneously, XPCS probes the microscopic dynamics at $q=2.61\,\mathrm{\AA}^{-1}$, close to the maximum of the FSDP, and is therefore sensitive to density fluctuations on the length scale associated with medium-range structural correlations~\cite{Ryu2025}. The microscopic dynamics are quantified by the two-time intensity autocorrelation function
\begin{equation}
    g_2(t_1,t_2) =
    \frac{\langle I(t_1)I(t_2)\rangle_\mathrm{p}}
    {\langle I(t_1)\rangle_\mathrm{p}\langle I(t_2)\rangle_\mathrm{p}},
    \label{trc}
\end{equation}
where $\langle...\rangle_\mathrm{p}$ denotes averaging over detector pixels within the selected $q$-range. The intensity correlation function is related to the normalized intermediate scattering function $f(q,t_1,t_2)$ through the Siegert relation,
\begin{equation}
    g_2(t_1,t_2)-1 = A\,|f(q,t_1,t_2)|^2,
\end{equation}
where $A$ denotes the experimental speckle contrast~\cite{Sutton2007}. The intermediate scattering function probes density-density correlations at the chosen wave vector and thus provides a direct measure of the microscopic structural relaxation.

The evolution of the microscopic dynamics following four representative temperature jumps is shown in Fig.~\ref{fig1}. The time-resolved autocorrelation functions are plotted as a function of the lag time $t_2-t_1$, with the color indicating the waiting time $t_1$ after the jump, where darker colors correspond to larger $t_1$. We define $t_1=0$ as the time at which the sample reaches the final temperature $T_\infty\pm0.1\,$K. The figure contains two temperature down-jumps [Fig.~\ref{fig1}(a,c)] and two up-jumps [Fig.~\ref{fig1}(b,d)], with the two larger jumps shown in the upper panels.

Before each jump, the sample is equilibrated at $T_0$, where it exhibits a stretched-exponential autocorrelation function characteristic of the equilibrium supercooled liquid, shown by the black dashed curves (equilibrium autocorrelation functions for a broad range of temperatures are shown in the SI). Following the jump to $T_\infty$, the relaxation changes markedly: the earliest autocorrelation functions exhibit strongly compressed-exponential behavior with $\beta>1$ (see red curve in panel (d)). This behavior is observed for up- and down-jumps independent of their magnitude and even for the smallest temperature change investigated in this study of $\Delta T=4\,$K (see SI). We note that these early autocorrelation functions in fact show a two-step decay, as a residual of the stretched-exponential $\alpha$-relaxation persists at short times ($<10\,$s), followed at longer times by a compressed relaxation~\cite{Shen2026,Frey2025}. As only the second process evolves during the equilibration within our temporal window, we focus on this process, as indicated by the fits shown in panel (d).

\begin{figure*}[t]
    \centering
       \includegraphics[width=0.99\textwidth]{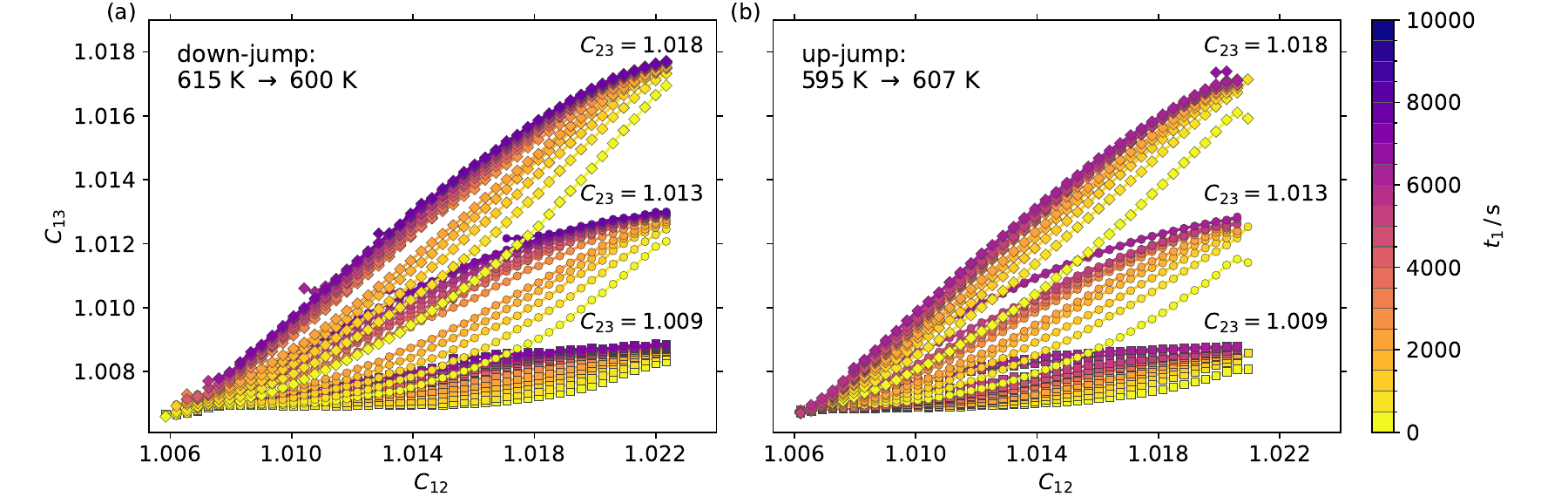}
    \caption{\label{fig2} Test of the triangular relation for two-time autocorrelation functions during aging after (a) a temperature down-jump from 615 K to 600 K and (b) an up-jump from 595 K to 607 K. The correlation $C_{13}$ is plotted as a function of $C_{12}$ for three fixed values of $C_{23}=1.009$, 1.013, and 1.018. Colors indicate the waiting time $t_1$ after the temperature jump. For a valid material-time description, $C_{13}$ should be uniquely determined by $C_{12}$ and $C_{23}$ and therefore independent of $t_1$. The systematic dependence on $t_1$ observed for both temperature jumps demonstrates a violation of the triangular relation.}
\end{figure*}

During subsequent aging at fixed $T_\infty$, the relaxation time increases and the shape of the correlation function evolves continuously from compressed towards stretched exponential. At long times, the correlation function becomes stationary and coincides with the equilibrium relaxation at $T_\infty$ (gray dashed curves).

Therefore we can conclude that compressed relaxation does not require driving the system far from equilibrium as done in previous metallic-glass XPCS studies, but emerges already following small temperature jumps from an equilibrated state. Importantly, the relaxation shape evolves continuously with age from compressed, $\beta>1$, to the stretched-exponential equilibrium shape, $\beta=0.65$. Compressed relaxation is therefore a transient feature of the non-equilibrium state rather than a stationary relaxation shape at that temperature. Resolving the complete evolution to equilibrium is enabled by the controlled small-temperature-jump protocols employed here. In previous XPCS studies, the lower temperatures and larger departures from equilibrium led to equilibration times that would have been much longer than the experimental observation window, preventing the full crossover to equilibrium from being resolved.

A second striking feature is that the relaxation times during aging do not simply evolve between their initial and final equilibrium values, $\tau(T_0)$ and $\tau(T_\infty)$. This is particularly evident following the up-jumps: immediately after the jump, the dynamics dramatically accelerates by more than one order of magnitude and the autocorrelation functions decay faster than the equilibrium correlation function at $T_\infty$, corresponding to relaxation times shorter than $\tau(T_\infty)$. During subsequent aging at fixed $T_\infty$, the dynamics then slows down towards the equilibrium value, rather than exhibiting the expected gradual speedup (see SI for an illustration). Thus, driving the system out of equilibrium produces a transient acceleration of the microscopic dynamics that exceeds the equilibrium relaxation rate at the final temperature.

The continuous evolution from compressed to stretched exponential relaxation raises the question of whether the changing correlation shape can nevertheless be represented by an evolving material-time variable. As discussed above, the change in shape alone is not sufficient to establish a breakdown of the material-time formalism, since a non-uniform material-time variable can modify the apparent relaxation shape as a function of laboratory time. We therefore test the triangular relation, Eq.~(\ref{eq:triangular-relation}), directly from the measured two-time correlations, without explicitly constructing a material-time variable.

For a total of $120\times 10^6$ time triplets $t_1<t_2<t_3$, we evaluate the pairwise autocorrelation values $C_{12}\equiv g_2(t_1,t_2)$, $C_{13}\equiv g_2(t_1,t_3)$, and $C_{23}\equiv g_2(t_2,t_3)$. The triplets are grouped according to their values of $C_{12}$, $C_{23}$, and their age $t_1$, and the corresponding $C_{13}$ values are averaged within each group. Figure~\ref{fig2} shows $C_{13}$ as a function of $C_{12}$ at three fixed values of $C_{23}$ for the large down-jump [panel (a)] and up-jump [panel (b)], with color indicating $t_1$. If the correlations obey the triangular relation, $C_{13}$ is uniquely determined by $C_{12}$ and $C_{23}$ and must therefore be independent of $t_1$. Instead, we observe a pronounced and systematic $t_1$-dependence for both temperature jumps and for all three values of $C_{23}$. The microscopic correlations therefore violate the triangular relation, demonstrating that their evolution during aging cannot be described in terms of a material time. 

\begin{figure}
    \centering
       \includegraphics[width=0.49\textwidth]{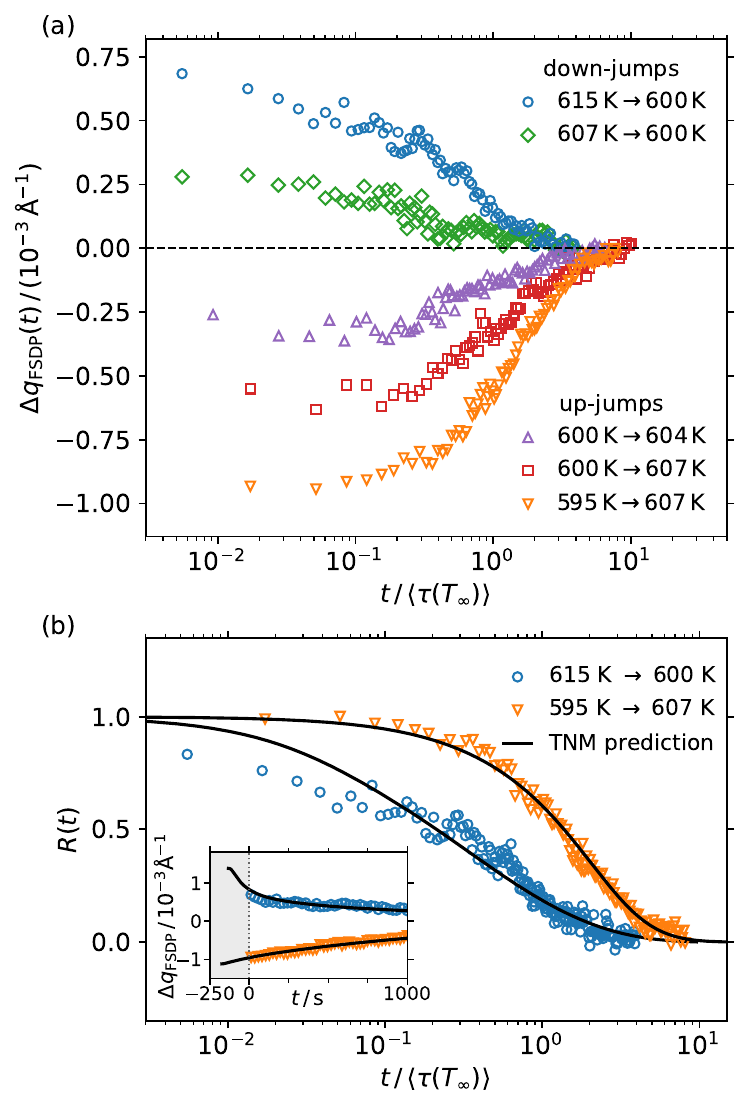}
    \caption{\label{fig3} Aging of the average structure following temperature up- and down-jumps. (a) Evolution of the FSDP position $q_\mathrm{FSDP}$ relative to its equilibrium value at the final temperature, plotted as a function of time normalized by the average equilibrium relaxation time $\langle\tau(T_\infty)\rangle$. (b) Normalized structural relaxation $R(t)$ for the largest down- and up-jumps. Solid black lines show the corresponding TNM predictions using a common parameter set. The inset shows the evolution of $\Delta q_\mathrm{FSDP}(t)$ around $t=0$, including the TNM prediction during the finite-duration temperature jump before $T_\infty$ is reached ($t<0$).}
\end{figure}

This raises the question of whether the breakdown of material-time behavior for the microscopic dynamics also extends to the evolution of the average structure. We therefore consider the temporal evolution of the FSDP position $q_\mathrm{FSDP}(t)$, which corresponds to the length scale the dynamics are probed at. Although $q_\mathrm{FSDP}(t)$ reflects the microscopic structure of the MG at MRO length scales, it represents an ensemble-averaged structural observable. Provided that the shape of the FSDP does not change qualitatively during aging, shifts of $q_\mathrm{FSDP}$ approximately reflect changes in the average density, with $\rho\propto q_\mathrm{FSDP}^3$~\cite{Yavari2005,Giordano2016}.

Figure~\ref{fig3}a shows the evolution of $q_\mathrm{FSDP}(t)$ following temperature jumps of different magnitude and direction. To compare jumps ending at different final temperatures, we consider the deviation from the equilibrium value at $T_\infty$,
\begin{equation}
    \Delta q_\mathrm{FSDP}(t) = q^\mathrm{eq}_\mathrm{FSDP}(T_\infty) - q_\mathrm{FSDP}(t),
\end{equation}
and normalize time by the equilibrium mean relaxation time $\langle\tau(T_\infty)\rangle$ at the respective final temperature. This scaling removes the trivial temperature dependence of the equilibrium relaxation rate and allows the aging following the different jumps to be compared on a common timescale.

In these reduced units, $q_\mathrm{FSDP}(t)$ reaches its equilibrium value after approximately $3\langle\tau(T_\infty)\rangle$ for the down-jumps and $7\langle\tau(T_\infty)\rangle$ for the up-jumps. Thus, equilibration of the average structure occurs on the scale of only a few equilibrium relaxation times at the final temperature. Aging timescales of the same order have been observed for molecular glass formers when expressed relative to their equilibrium relaxation times~\cite{Richert2023,Henot2024}, although these studies predominantly probe rotational rather than atomic density relaxation.

Beyond this difference in equilibration time, Fig.~\ref{fig3}a reveals an asymmetry between temperature up- and down-jumps. Following a down-jump, the structure initially evolves rapidly and subsequently slows as equilibrium is approached, whereas following an up-jump the initial evolution is slower and accelerates during aging. The down-jumps consequently reach equilibrium substantially earlier than the corresponding up-jumps. This asymmetry is highlighted in Fig.~\ref{fig3}b, where the two largest jumps are compared using a normalized representation of the structural evolution, $R(t)$, as defined below. Such an asymmetry of approach is a characteristic feature of material-time aging and results naturally from the opposite evolution of the internal clock rate following up- and down-jumps.

To test quantitatively whether the structural evolution is consistent with this picture, we describe both jumps using the TNM model. In brief, we use the measured temperature history $T(t)$ together with the independently determined equilibrium relaxation time $\langle\tau(T)\rangle$ to predict the evolution of the fictive temperature $T_\mathrm{f}(t)$ that characterizes the structural state of the glass. $T_\mathrm{f}(t)$ is then converted into a TNM prediction for the evolution of the FSDP position $q_\mathrm{FSDP}^\mathrm{TNM}(t)$ using the independently measured equilibrium structure-temperature relation $q^\mathrm{eq}_\mathrm{FSDP}(T)$. For more details, see the Materials and Methods section and the SI.

Because the temperature jump is not instantaneous, some aging already occurs at $t<0$ while the sample temperature is still being changed and has not yet reached $T_\infty$. The normalization in Fig.~\ref{fig3}b is performed with respect to the structural state predicted by the TNM model at $t=0$. Specifically, we define
\begin{equation}
    R(t)= \frac{\Delta q_\mathrm{FSDP}(t)}{\Delta q_\mathrm{FSDP}^\mathrm{TNM}(0)},
\end{equation}
where $\Delta q_\mathrm{FSDP}^\mathrm{TNM}(0)$ is the departure from equilibrium predicted by the TNM model at the time the sample first reaches $T_\infty$. This normalization allows the relaxation shapes of the up- and down-jumps to be compared directly while accounting for the different amounts of structural evolution that occured at $t<0$.

Using a common set of TNM parameters for both jumps, the model provides a satisfactory description of $q_\mathrm{FSDP}(t)$, as shown by the black curves in Fig.~\ref{fig3}b. In particular, it reproduces the pronounced asymmetry between the up- and down-jumps, as well as the time required to reach equilibrium. Importantly, the TNM calculation is based on the experimentally measured sample temperature history $T(t)$ at $t<0$, instead of using the first measured value as an initial condition. The inset of Fig.~\ref{fig3}b shows the corresponding prediction for $q_\mathrm{FSDP}(t)$ during this interval at $t<0$ (grey area). Although no experimental data are available at $t<0$, the predicted value of $q_\mathrm{FSDP}(t=0)$ agrees well with the first experimental data points. The TNM description therefore also accounts for the structural evolution occurring while the sample temperature is still being changed at $t<0$.

The characteristic asymmetry of approach and its quantitative description by the TNM model show that the evolution of the average structure is consistent with the material-time formalism. This stands in marked contrast to the microscopic two-time correlations, which violate the triangular relation and, therefore, cannot be described in terms of a material time.

Although the microscopic dynamics and the evolution of the average structure differ fundamentally with respect to their compatibility with a material-time description, both evolve continuously during the aging process. This raises the question of how the transient change from compressed to stretched relaxation is related to the approach of the average structure towards equilibrium.

\begin{figure}
    \centering
       \includegraphics[width=0.49\textwidth]{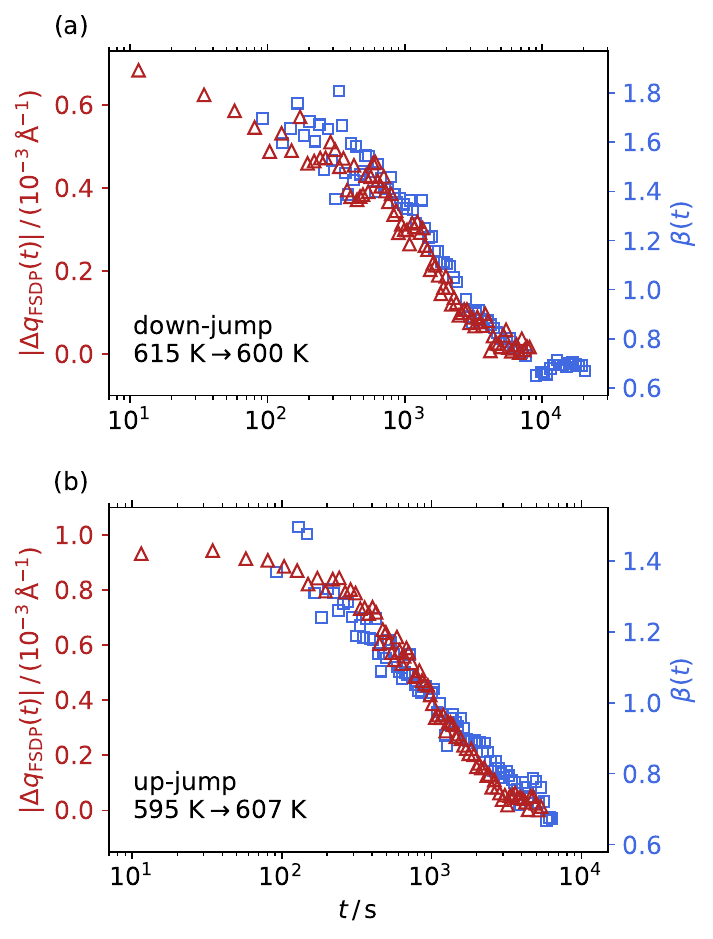}
    \caption{\label{fig4} Evolution of the deviation from equilibrium of the FSDP maximum, $|q_\mathrm{FSDP}(t)-q_\mathrm{FSDP}^{\mathrm{eq}}(T_\infty)|$ (red, left axes), and the KWW stretching exponent $\beta(t)$ (blue, right axes) during aging after (a) a temperature down-jump from 615 K to 600 K and (b) an up-jump from 595 K to 607 K. In both cases, $\beta$ approaches its equilibrium value on approximately the same timescale on which $q_\mathrm{FSDP}$ reaches equilibrium.}
\end{figure}

To address this question, Fig.~\ref{fig4} compares the evolution of the stretching exponent $\beta(t)$ with the deviation of the structure from equilibrium, $|\Delta q_\mathrm{FSDP}(t)|$, for representative up- and down-jumps. Here, $\beta(t)$ is obtained by fitting the final decay of the time-resolved autocorrelation functions using the Kohlrausch-William-Watts (KWW) function
\begin{equation}
    g_2(t_1,t_2) = c\,\exp\left[ -2\left(\frac{t_2 - t_1}{\tau}\right)^\beta\right] +b,
    \label{equ:KWW}
\end{equation}
where the parameter $\beta$ distinguishes stretched ($\beta<1$) and compressed ($\beta>1$) relaxation. The factor of two in the exponential function arises from using intensity instead of electric-field autocorrelation functions. Two exemplary fits by Eq.~(\ref{equ:KWW}) are included in Fig.~\ref{fig1}d.

As evident from Fig.~\ref{fig4}, the stretching exponent $\beta$ and the evolution of the FSDP position follow a similar temporal evolution and approach their respective equilibrium values simultaneously. In particular, the compressed relaxation shape indicated by enhanced values of $\beta$ disappears as soon as $q_\mathrm{FSDP}$ reaches its equilibrium value, implying that the stretched equilibrium relaxation, $\beta_\mathrm{eq}\simeq0.65$, is recovered once structural equilibration is complete. Thus, although the microscopic correlations cannot be described by a material time, the transient change of their relaxation shape remains closely linked to the non-equilibrium structural state of the glass.

\section*{Discussion}

Our results reveal a clear difference between physical aging at the level of the average structure and at the level of microscopic fluctuations. The evolution of $q_\mathrm{FSDP}$ displays the characteristic asymmetry between temperature up- and down-jumps expected for material-time aging and is well described by the TNM model using a common parameter set. By contrast, the microscopic two-time autocorrelation functions violate the triangular relation and therefore cannot be rationalized in terms of a material time. Thus, a material-time description can be consistent with the evolution of the average structure without providing a complete description of the microscopic dynamics during aging.

At the same time, our measurements provide new insight into the conditions under which compressed-exponential relaxation occurs in MGs. We observe compressed autocorrelation functions even after temperature changes as small as $\Delta T=4\,$K from an equilibrated supercooled liquid. Compressed relaxation is therefore not restricted to deep glassy states or to samples prepared far from equilibrium. The emergence of strongly compressed autocorrelation functions even after small perturbations from equilibrium, irrespective of the direction of the temperature change, goes beyond previous interpretations of compressed-exponential relaxation based on a temperature-dependent competition between thermal and plastic relaxation~\cite{Rodriguez-Lopez2023} and calls for further investigation.

At fixed $T_\infty$, the relaxation shape changes continuously with age from compressed- to stretched-exponential. This evolution occurs on the same timescale on which the average structure approaches equilibrium. While this does not imply a unique relation between $\beta$ and $q_\mathrm{FSDP}$, it shows that the relaxation shape is closely linked to the nature of the structure at MRO length scales, as suggested also by previous dynamical studies in equilibrium supercooled liquids~\cite{Neuber2022}. 

The observation that compressed-exponential relaxation vanishes once the structure has equilibrated is consistent with recent interpretations of compressed relaxation in MGs in terms of stress-mediated dynamics associated with medium-range structural correlations~\cite{Frey2025,Shen2026,Ryu2025}, potentially driven by thermally activated plastic events~\cite{Ferrero2014,Rodriguez-Lopez2023}. In this picture, vitrification freezes the MRO into a non-equilibrium configuration, leading to internal stresses, whose relaxation generates additional atomic displacements that strongly contribute to the decay of density correlations at the FSDP. This stress-mediated contribution is thus expected to disappear once equilibrium is restored, offering a physical explanation for the observed breakdown of material-time scaling at the microscopic level. The material-time concept assumes that aging is governed by the same underlying relaxation processes as the equilibrium dynamics, with only the internal clock rate changing. The additional process associated with compressed relaxation, which is present only out of equilibrium, therefore cannot be captured by such a reparameterization of time.


Our results further suggest that the enhanced microscopic dynamics associated with the compressed-exponential relaxation does not directly translate into a faster evolution of the average structure. In particular, although the microscopic relaxation after a temperature up-jump is faster than in equilibrium at $T_\infty$, the average structure reaches equilibrium only after several equilibrium relaxation times. Stress-mediated dynamics therefore produces substantial microscopic rearrangements without correspondingly accelerating the evolution of the average structure. This decoupling between structural evolution and dynamics supports a picture of non-density-driven atomic motion in the glassy state, consistent with previous studies of MGs under high pressure compressions~\cite{cornet2023,ronca2025}.

At the level of the evolution of average observables, the aging observed here is similar to that reported for molecular glass formers. When time is expressed in units of the equilibrium relaxation time $\langle\tau(T_\infty)\rangle$, the average structure reaches equilibrium after only a few relaxation times, comparable to aging timescales reported for molecular glasses~\cite{Richert2023,Henot2024}. Together with the asymmetry between up- and down-jumps and the successful TNM description, this suggests a similar phenomenology of physical aging across chemically different glass formers. At the microscopic level, however, the picture is different. Microscopic reorientation dynamics measured by depolarized dynamic light scattering in molecular glasses has been found to satisfy the triangular relation and thus to be consistent with a material-time description~\cite{Boehmer2024}, whereas the density-density fluctuations measured here violate the triangular relation. It is not yet clear whether this violation is specific to MGs or instead depends on the observable and length scale being probed. The present XPCS measurements probe density fluctuations at the FSDP and are therefore sensitive to atomic motion associated with medium-range structural correlations. Recent XPCS studies have shown that compressed dynamics in MGs are particularly linked to these length scales, while stretched exponential decays are recovered when probing larger scales, even in the deep glass state~\cite{Shen2026}. By contrast, depolarized light scattering probes molecular reorientation dynamics. Measurements of aging in MGs after small temperature jumps at other wave vectors than those corresponding to the FSDP could help determine whether the violation of material time observed here is specifically connected to medium-range-order dynamics or is a more general feature of aging in MGs.

From atomistic simulations, material-time behavior has been reported for potential-energy autocorrelation functions~\cite{Boehmer2024} and incoherent intermediate scattering functions~\cite{Amari2026}, in contrast to the violation observed here for collective density-density fluctuations at the FSDP. Whether this difference originates from the observable, the studied material, or the constant-volume conditions used in these simulations remains an open question. Importantly, our finding that compressed-exponential relaxation emerges after small temperature changes close to equilibrium suggests that such dynamics may be accessible in simulations without the need to reach very low temperatures.

In summary, we performed controlled temperature-jump experiments on a MG that reveal a clear difference between the evolution of the average structure and the microscopic dynamics. While the evolution of the FSDP position is consistent with a material-time description and is described by the TNM model, the microscopic two-time autocorrelation functions strongly violate the triangular relation and therefore cannot be rationalized in terms of a material time. At the same time, we find that compressed-exponential relaxation is observed even after small temperature jumps and gradually disappears as the average structure approaches equilibrium, consistent with an origin in stress-mediated relaxation associated with volumetric frustration. This additional non-equilibrium relaxation, absent in equilibrium, provides a possible physical explanation for the microscopic breakdown of material-time scaling, since the latter assumes that aging can be described as a reparameterization of the equilibrium dynamics. Our results imply that the validity of a material-time description can depend on the level at which aging is probed. A material-time variable can provide an effective description of the evolution of the average structure even when the underlying microscopic dynamics does not obey material-time scaling. 

\section*{Acknowledgments}
We gratefully acknowledge the European Synchrotron Radiation Facility (ESRF) for provision of synchrotron radiation facilities under proposal HC5966 on beamline ID10-COH. We thank Marco Cammarata and Yuriy Chushkin for support regarding data analysis during the beamtime. Furthermore, we would like to thank Jeppe C. Dyre for fruitful discussions and valuable suggestions for improving the manuscript, as well as Linnea Heitmeier for helpful discussions concerning possible simulation approaches.

\section*{Author contributions}
T.B.: Conceptualization, Methodology, Formal analysis, Investigation, Data curation, Software, Visualization, Project administration, Writing – original draft, Writing – review \& editing. J.S., J.P.G., R.Z., F.Z., F.Y., E.P.: Investigation, Writing – review \& editing. B.R.: Conceptualization, Methodology, Investigation, Resources, Writing – review \& editing. Th.B.: Methodology, Investigation, Resources, Supervision, Writing – review \& editing. All authors discussed the results and contributed to the final manuscript.

\section*{Data availability}
The raw XRD and XPCS data generated in this study are archived at the European Synchrotron Radiation Facility (ESRF). Processed data supporting the findings of this study are available from the corresponding authors upon reasonable request.

\section*{Materials and Methods}

\subsection*{Sample preparation}
The MG sample with a composition of Vit4 (Zr$_{46.8}$Ti$_{8.2}$Cu$_{7.5}$Ni$_{10}$Be$_{27.5}$, in atomic percent) was synthesized from elements with a minimum purity of 99.99\%. The Vit4 alloy was arc melted and subsequently processed into a ribbon by melt spinning under a high-purity argon atmosphere, resulting in a thickness of 30 $\pm$ 5 $\mu$m. The obtained ribbon was polished to a thickness of 10 $\mu$m to achieve a satisfactory speckle contrast for the XPCS experiments. The sample was mounted in a resistively heated, PID-controlled furnace designed for transmission X-ray scattering measurements.

The experiment was performed at the ID10-COH beamline of the European Synchrotron Radiation Facility (ESRF) using an incident X-ray energy of 15.875\,keV. The beam was focused to a cross section of approximately $7.8\times3.9\,\mu\mathrm{m}^2$ at the sample position. X-ray diffraction (XRD) measurements using a Si PILATUS 300k detector and X-ray photon correlation spectroscopy (XPCS) measurements using a CdTe Eiger2 4M detector were performed simultaneously, allowing the temporal evolution of the average structure and microscopic dynamics to be monitored under identical thermal conditions.

\subsection*{Temperature-jump protocol}
Prior to each temperature-jump experiment, the sample was equilibrated at an initial temperature $T_0$ in the supercooled-liquid regime by waiting for at least ten times the equilibrium relaxation time $\tau_\mathrm{eq}(T_0)$. We validated that equilibrium was reached by confirming that structural and dynamical observables showed no further systematic temporal evolution within the experimental resolution. The sample temperature was then changed to a final temperature $T_\infty$ as quickly as possible (see SI for selected data of the temperature evolution), producing temperature jumps $\Delta T=T_\infty-T_0$ of either sign. The XRD and XPCS measurements started as soon as the sample temperature reached $T_\infty\pm 0.1\,$K, which we define as $t=0$. The aging process was then monitored until the structural and dynamical observables reached their equilibrium values at $T_\infty$. To avoid impractically large data sets, each measurements was split into several batches (between three and nine batches per jump). The complete measured temperature history $T(t)$ was recorded and used in the subsequent analysis and modeling.

\subsection*{Time-resolved X-ray diffraction}
The time-dependent scattering intensity $I(q,t)$ was recorded using a Pilatus 300K detector over a momentum-transfer range of $q=1.4$--$4.4\,\mathrm{\AA}^{-1}$, covering the FSDP of Vit4. The position of the FSDP, $q_{\mathrm{FSDP}}(t)$, was determined by fitting a Lorentzian function,
\begin{equation}
I(q)=A\,\frac{\Gamma^2}{\Gamma^2+(q-q_\mathrm{FSDP})^2},
\end{equation}
to the FSDP over the range $2.46-2.76\,\mathrm{\AA}^{-1}$, with $q_\mathrm{FSDP}$, $A$, and $\Gamma$ as free parameters.

Equilibrium measurements at different temperatures were conducted to determine the equilibrium
relation $q_{\mathrm{FSDP}}^{\mathrm{eq}}(T)$, which provides the reference values for the
temperature-jump experiments.

Over the course of four days of beam time, the calibration of the $q$-vectors was subject to a slight linear drift with respect to the equilibrium measurements conducted within the first few hours of the experiment. All values of $q_{\mathrm{FSDP}}(t)$ probed during the aging experiments were corrected for this offset. Due to the slow evolution of the offset, its magnitude can be considered constant over the course of a single aging experiment, while it needs to be taken into account when comparing experiments conducted on different days.

\subsection*{X-ray photon correlation spectroscopy}

XPCS measurements were performed using an Eiger 4M CdTe detector positioned at a sample--detector distance of 4.96\,m, at $q=2.61\,\mathrm{\AA}^{-1}$, corresponding approximately to the maximum of the FSDP, with time-resolution 0.02\,s. 

To improve the signal-to-noise ratio of the time-resolved autocorrelation functions, the two-time correlation function was averaged over 16\,s intervals of the waiting time $t_1$. This interval is short compared with the characteristic timescale of the aging process, such that the evolution of the system within a single averaging window is negligible, which we confirmed by comparing the results to data averaged over shorter $t_1$-intervals. The resulting $t_1$-resolved autocorrelation functions therefore retain the temporal evolution of the dynamics while providing a higher signal-to-noise ratio.

In addition, the lag-time axis $\Delta t=t_2-t_1$ was logarithmically binned. This reduces statistical noise at long lag times, where fewer independent intensity pairs contribute to the correlation function, while preserving a fine time resolution at short lag times.

\subsection*{Triangular relation}

To test the triangular relation, for each measured batch we construct triplets of times $t_1<t_2<t_3$ from the measured two-time autocorrelation matrix. Starting times $t_1$ are sampled every 200 frames ($4.6\,\mathrm{s}$), while the intervals $t_2-t_1$ and $t_3-t_2$ are independently chosen from logarithmically spaced values between $5$ and $1000\,\mathrm{s}$. For each valid triplet, we determine the three pairwise correlations $C_{12}\equiv g_2(t_1,t_2)$, $C_{23}\equiv g_2(t_2,t_3)$, and $C_{13}\equiv g_2(t_1,t_3)$.

To reduce statistical noise, each correlation value is locally averaged in the two-time correlation matrix over neighboring starting times within $\pm200$ frames ($\pm4.6\,\mathrm{s}$) and neighboring lag times within approximately $\pm2.5\,\mathrm{s}$. The resulting triplets from the individual batches are combined and binned according to the age $t_1$ and the correlation values $C_{12}$ and $C_{23}$. We use equally spaced $t_1$ bins between $0$ and $10000\,\mathrm{s}$ and correlation bins between $1.0055$ and $1.0227$. For each $(t_1,C_{12},C_{23})$ bin, the corresponding values of $C_{13}$ are averaged, and their standard deviation and number of contributing triplets are considered for determining the statistical error.
\\
\subsection*{TNM analysis}
The evolution of $q_\mathrm{FSDP}(t)$ was modeled using the Tool--Narayanaswamy--Moynihan (TNM) framework. The model describes the structural state by a fictive temperature $T_\mathrm{f}(t)$ and uses the phenomenological material-time clock rate
\begin{equation}
\frac{\mathrm{d}\xi}{\mathrm{d}t}=\left[\tau_\mathrm{eq}\left(T(t)\right)^x~\tau_\mathrm{eq}\left(T_\mathrm{f}(t)\right)^{1-x}\right]^{-1}.
\end{equation}
where $x$ determines the relative contributions of the actual and fictive temperatures to the aging rate, and $\tau_\mathrm{eq}(T)$ is the independently measured equilibrium relaxation time. For an arbitrary temperature history, the fictive temperature evolves according to
\begin{equation}
T_\mathrm{f}(t)=T(t)-\int_{t_\mathrm{i}}^{t}M\left[\xi(t)-\xi(t')\right]\frac{\mathrm{d}T(t')}{\mathrm{d}t'},\mathrm{d}t',
\end{equation}
with $T_\mathrm{f}(t_\mathrm{i})=T(t_\mathrm{i})=T_0$ in the initial equilibrium state. We assume a stretched-exponential memory kernel with the same stretching exponent, $\beta=0.65$, as the equilibrium structural relaxation measured by XPCS,
\begin{equation}
M(\Delta\xi)=\exp\left[-\left(\frac{\Delta\xi}{\xi_0}\right)^{0.65}\right].
\end{equation}
A common parameter set, $x=0$ and $\xi_0=3.75$, is used to describe the temperature up- and down-jumps. Here, $x=0$ implies that the relaxation time relevant for aging is only determined by the fictive and not by the actual temperature. The calculated fictive temperature is converted into a prediction for $q_\mathrm{FSDP}(t)$ using the independently measured equilibrium structure-temperature relation $q_\mathrm{FSDP}^\mathrm{eq}(T)$ and assuming
\begin{equation}
q_\mathrm{FSDP}^\mathrm{TNM}(t)=q_\mathrm{FSDP}^\mathrm{eq}\left[T_\mathrm{f}(t)\right].
\end{equation}
The experimentally measured temperature history $T(t)$, including the finite-duration temperature change at $t<0$, is used directly in the calculation. Consequently, the structural state at $t=0$ is predicted by the model rather than set by the first measured value of $q_\mathrm{FSDP}$. The determination of $\tau_\mathrm{eq}(T)$ and $q_\mathrm{FSDP}^\mathrm{eq}(T)$, as well as further details of the calculation, are provided in the Supplemental Material.

\bibliography{bib_metallic.bib}
\end{document}